\documentclass[twocolumn]{revtex4}
\usepackage{mathbbold}
\usepackage{amsfonts}
\usepackage{amsmath}
\usepackage{amssymb}
\usepackage{charter}
\usepackage{graphicx}

\begin{document}

\title{Noiselessly amplified thermal states and after multi-photon addition
or subtraction}
\author{Xue-feng Zhan and Xue-xiang Xu$^{\dag }$ }
\affiliation{College of Physics and Communication Electronics, Jiangxi Normal University,
Nanchang 330022, China\\
$^{\dag }$xuxuexiang@jxnu.edu.cn}

\begin{abstract}
In this paper, we introduce a noiselessly amplified thermal state (ATS), by
operating the noiseless amplification operator ($g^{\hat{n}}$)\ on the
thermal state (TS) with corresponding mean photon number (MPN) $\bar{n}$.
Actually, the ATS is an new TS with MPN $\bar{N}=g^{2}\bar{n}/[1-\bar{n}%
\left( g^{2}-1\right) ]$. Furthermore, we introduce photon-added-ATS (PAATS)
and photon-subtracted-ATS (PSATS) by operating $m$-photon addition ($\hat{a}%
^{\dag m}$) and $m$-photon subtraction ($\hat{a}^{m}$) on the ATS,
respectively. We study photon number distributions (PNDs), purities, and
Wigner functions (WFs) for all these states.

\end{abstract}

\maketitle

\section{Introduction}

Quantum theory is originated from Planck's discovery of the radiation law%
\cite{1,2}. Often, those electromagnetic radiations emitted by a hot body
with a certain temperature were called as thermal lights\cite{3}. One of
them, namely thermal state (TS), can be expressed in terms of Fock states%
\begin{equation}
\rho _{th}\left( \bar{n}\right) =\sum_{n=0}^{\infty }\frac{\bar{n}^{n}}{(%
\bar{n}+1)^{n+1}}\left\vert n\right\rangle \left\langle n\right\vert ,
\label{a1}
\end{equation}%
where $\bar{n}$\ denotes its mean photon number (MPN)\cite{4}. Usually, the
TS plays a significant role in statistical physics\cite{5}. However, this TS
is an Gaussian state, which lack some desirable properties (e.g., Wigner
negativity) for quantum supremacy in various applications\cite{6,7}.

Photon addition (by operating creation operator $\hat{a}^{\dag }$) and
photon subtraction (by operating annihilation operator $\hat{a}$), as the
basic tools, have been used to generate non-Gaussian quantum states\cite{8}
with some desirable properties. Recently, Guerrini et al. called these
states as photon-varied quantum states\cite{9}. Barnett et al. presented a
detailed analysis of the photon statistics for photon-subtracted and
photon-added states\cite{10}. Based on the TS, the reseachers have
introduced many quantum states and done many theoretical and experimental
works\cite{11,12,13}. Zavatta et al. demonstrated the bosonic communication
relation in experiment\cite{14} by applying $\hat{a}\hat{a}^{\dag }-e^{i\phi
}\hat{a}^{\dag }\hat{a}$ on the TS. Zavatta et al. reported the generation
and the analysis of single-photon-added or single-photon-subtracted TSs\cite%
{15} by applying $\hat{a}^{\dag }$ or $\hat{a}$ on the TS. Many groups have
introduced and studied the multiphoton-subtracted TSs by using $\hat{a}^{m}$
($m$ is an integer) \cite{16,17,18,19,20}. Chatterjee compared the
nonclassicality of the added-then-subtraced and subtracted-then-added TS by
using $\hat{a}^{q}\hat{a}^{\dag p}$ or $\hat{a}^{\dag p}\hat{a}^{q}$\ ($p,q$
are integers)\cite{21}. Deepak and Chatterjee reported a study related with
a coherent superposition TS by using $s\hat{a}\hat{a}^{\dag }+t\hat{a}^{\dag
}\hat{a}$ \cite{22}. Our group has introduced photon
annihilation-then-creation TS and photon creation-then-annihilation TS by
using $\left( \hat{a}^{\dag }\hat{a}\right) ^{m}$ and $\left( \hat{a}\hat{a}%
^{\dag }\right) ^{m}$\cite{23}. Our group has also introduced a kind of
amplified TS by using $\left( g-1\right) \hat{a}^{\dag }\hat{a}+1$\cite{24},
where $g>1$\ is the amplification gain.

In recent decades, studies on the noiseless quantum amplifications have
attracted significant attention of researchers\cite{25,26,27,28,29,30}. As
we know, noise is unavoidably added to the amplified signal in any
deterministic linear amplification. Observing the results in the
hypothetical noiseless amplification of coherent state (i.e., $\left\vert
\alpha \right\rangle \rightarrow \left\vert g\alpha \right\rangle $), one
can infer that it is not a deterministic physical operation (i.e., a
trace-preserving positive map)\cite{31}. In order to overcome the limit
of deterministic amplification, one can resort to probabilistic amplification%
\cite{32}, which can be approximate the conditional operation $g^{\hat{n}}$ (%
$g>1$ and $\hat{n}=\hat{a}^{\dag }\hat{a}$). Thus, the input coherent state $%
\left\vert \alpha \right\rangle $\ can be mapped into an amplified coherent
state $\left\vert g\alpha \right\rangle $, i.e., $g^{\hat{n}}\left\vert
\alpha \right\rangle =e^{(g^{2}-1)\left\vert \alpha \right\vert
^{2}/2}\left\vert g\alpha \right\rangle $ \cite{33}. However, this operation
is still unphysical because no device realizing this transformation can be
implemented perfectly\cite{34}. The idea on the heralded noiseless
amplification was theoretically proposed by Ralph and Lund\cite{35}.
Afterwards, some amplifications were experimentally demonstrated by several
groups\cite{36,37,38}.

In a recent work, we have studied multi-photon-addition amplified coherent
states by considering the effects of $g^{\hat{n}}$\ and $\hat{a}^{\dag m}$%
\cite{39}. However, we have never seen any reports on the combined effect
of $g^{\hat{n}}$ with $\hat{a}^{\dag m}$ or with $\hat{a}^{m}$ on the TS. So
in this paper, we shall study this problem and introduce related quantum
states. The paper is structured as follows: we introduce the ATS in Sec.2,
and then introduce the PAATS and the PSATS in Sec.3. In Sec.4-6, we study
their photon number distribution (PND), purity, Wigner function (WF),
respectively. Our conclusions are given in the last section.

\section{Introduction of amplified TS}

As shown in Fig.1(a), we introduce\ another amplified TS (ATS) by
considering the effect of $g^{\hat{n}}$ on the TS. Making simple analysis\
(see Appendix A), we prove that the operator $g^{\hat{n}}$ can map $\rho
_{th}\left( \bar{n}\right) $\ into a new TS $\rho _{th}\left( \bar{N}\right)
$, i.e.%
\begin{equation}
g^{\hat{n}}\text{: }\rho _{th}\left( \bar{n}\right) \Longrightarrow \rho
_{th}\left( \bar{N}\right) ,  \label{aa}
\end{equation}%
in the condition of $g^{2}\bar{n}/(\bar{n}+1)<1$ and with new MPN $\bar{N}$.
Here, we name this new TS as noiselessly amplified TS (ATS) yielding
\begin{equation}
\rho _{th}\left( \bar{N}\right) =\frac{1}{N_{0}}g^{\hat{n}}\rho _{th}\left(
\bar{n}\right) g^{\hat{n}},  \label{a2}
\end{equation}%
with $N_{0}=[1-\bar{n}\left( g^{2}-1\right) ]^{-1}$ and $\bar{N}=N_{0}g^{2}%
\bar{n}$. Obviously, if $g=1$, leading to $\bar{N}=\bar{n}$, then $\rho
_{th}\left( \bar{N}\right) $ will reduce to $\rho _{th}\left( \bar{n}\right)
$. But if $g^{2}\bar{n}/(\bar{n}+1)>1$, it is impossible to form a quantum
state.

It is necessary to emphasize that the parameters ($g$\ and $\bar{n}$) must
be chosen rationally because the operator $g^{\hat{n}}$\ is unbounded. In
Fig.1(b), we plot the possible regions with $\bar{N}>0$ and $\bar{N}<0$ in
the ($\bar{n}$, $g$) parameter space, where the boundary line is satisfied $%
g=\sqrt{(\bar{n}+1)/\bar{n}}$. We find that: (i) Only in the condition of $%
1<g<\sqrt{(\bar{n}+1)/\bar{n}}$, a physical state can be generated due to $%
\bar{N}>0$. (ii) But in the condition of $g>\sqrt{(\bar{n}+1)/\bar{n}}$, it
will be unphysical because of $\bar{N}<0$.

To further demonstrate Fig.1(b), we plot Fig.1(c) and Fig.1(d) by setting
several parameters. In Fig.1(c), we plot $\bar{N}$\ as a function of $g$\
with $\bar{n}=1.1$ and $\bar{n}=1.5$. There is a critical value $g_{c}=\sqrt{%
(\bar{n}+1)/\bar{n}}$ for each $\bar{n}$, with $\bar{N}>0$ for $1<g<g_{c}$
and $\bar{N}<0$ for $g>g_{c}$. In the critical value $g=g_{c}$, $\bar{N}$
will be limited to positive infinity or negative infinity. In Fig.1(d), we
plot $\bar{N}$ as a function of $\bar{n}$\ with $g=1.06$ and $\bar{n}=1.08$.
Similarly, there is a critical value $\bar{n}_{c}=1/\left( g^{2}-1\right) $
for each $g$, with $\bar{N}>0$ for $0<\bar{n}<\bar{n}_{c}$ and $\bar{N}<0$
for $\bar{n}>\bar{n}_{c}$. In the critical value $\bar{n}=\bar{n}_{c}$, $%
\bar{N}$ will be limited to positive infinity or negative infinity.

\begin{figure*}[tbp]
\label{Fig1} \centering\includegraphics[width=2.0\columnwidth]{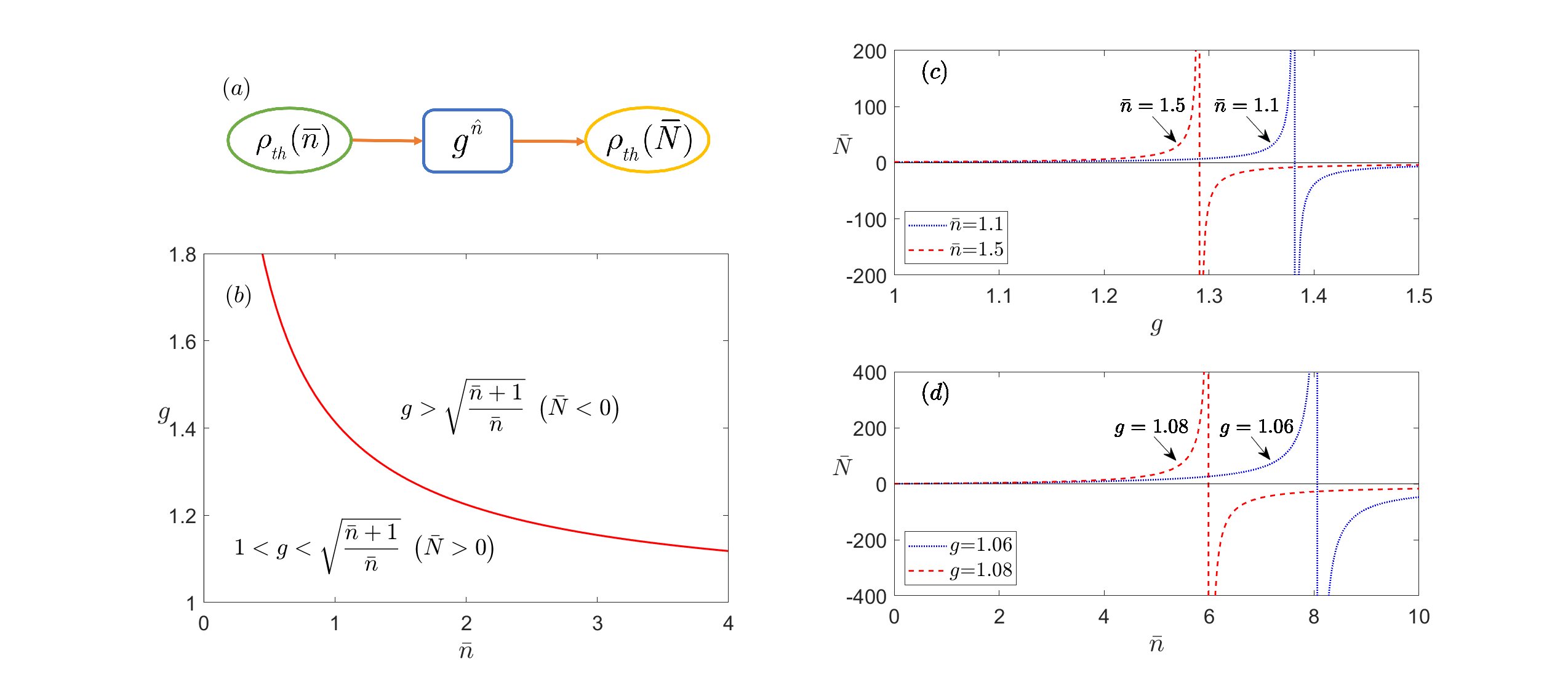}
\caption{(a) Conceptual scheme of generating ATS; (b) Regional plot of $\bar{%
N}>0$\ and $\bar{N}<0$\ in ($\bar{n},g$) space, where the boundary line is
constrained by $g=\protect\sqrt{(\bar{n}+1)/\bar{n}}$; (c) $\bar{N}$ versus $%
g$\ with $\bar{n}=1.1$, $1.5$ with respective $g_{c}\doteq 1.3817$, $1.2910$%
; (d) $\bar{N}$ versus $\bar{n}$\ with $g=1.06$, $1.08$ with respective $%
\bar{n}_{c}\doteq 8.0906$, $6.0096$.}
\end{figure*}

\section{Introduction of PAATS and PSATS}

Based on the ATS, we further introduce photon-added ATS (PAATS) and
photon-subtracted ATS (PSATS) by considering the effects of $\hat{a}^{\dag
m} $\ and $\hat{a}^{m}$.

\textit{PAATS:} As shown in Fig.2(a), applying $\hat{a}^{\dag m}$ then\ $g^{%
\hat{n}}$ (or $g^{\hat{n}}$ then\ $\hat{a}^{\dag m}$)\ on $\rho _{th}\left(
\bar{n}\right) $ yields the photon-added amplified TS (PAATS)
\begin{eqnarray}
\rho _{m+} &=&\frac{1}{N_{m+}^{(1)}}g^{\hat{n}}\hat{a}^{\dag m}\rho
_{th}\left( \bar{n}\right) \hat{a}^{m}g^{\hat{n}}  \notag \\
&=&\frac{1}{N_{m+}^{(2)}}\hat{a}^{\dag m}g^{\hat{n}}\rho _{th}\left( \bar{n}%
\right) g^{\hat{n}}\hat{a}^{m}  \notag \\
&=&\frac{1}{N_{m+}}\hat{a}^{\dag m}\rho _{th}\left( \bar{N}\right) \hat{a}%
^{m},  \label{a3}
\end{eqnarray}%
with the normalization coefficients $N_{m+}=m!\left( \bar{N}+1\right) ^{m}$,
$N_{m+}^{(1)}=g^{-2m}N_{0}N_{m+}$, and $N_{m+}^{(2)}=N_{0}N_{m+}$. In the
second step, we have used $g^{\hat{n}}\hat{a}^{\dag m}=g^{m}\hat{a}^{\dag
m}g^{\hat{n}}$. In particularly, $\rho _{0+}$\ is just $\rho _{th}\left(
\bar{N}\right) $.

\textit{PSATS:} As shown in Fig.2(b), applying $\hat{a}^{m}$ then\ $g^{\hat{n%
}}$ (or $g^{\hat{n}}$ then\ $\hat{a}^{m}$)\ on $\rho _{th}\left( \bar{n}%
\right) $ yields the photon-subtracted amplified TS (PSATS)
\begin{eqnarray}
\rho _{m-} &=&\frac{1}{N_{m-}^{(1)}}g^{\hat{n}}\hat{a}^{m}\rho _{th}\left(
\bar{n}\right) \hat{a}^{\dag m}g^{\hat{n}}  \notag \\
&=&\frac{1}{N_{m-}^{(2)}}\hat{a}^{m}g^{\hat{n}}\rho _{th}\left( \bar{n}%
\right) g^{\hat{n}}\hat{a}^{\dag m}  \notag \\
&=&\frac{1}{N_{m-}}\hat{a}^{m}\rho _{th}\left( \bar{N}\right) \hat{a}^{\dag
m},  \label{a4}
\end{eqnarray}%
with the normalization coefficients $N_{m-}=m!\bar{N}^{m}$, $%
N_{m-}^{(1)}=g^{2m}N_{0}N_{m-}$, and $N_{m-}^{(2)}=N_{0}N_{m-}$. In the
second step, we have used $g^{\hat{n}}\hat{a}^{m}=g^{-m}\hat{a}^{m}g^{\hat{n}%
}$. In particularly, $\rho _{0-}$\ is just $\rho _{th}\left( \bar{N}\right) $%
.

\begin{figure}[tbp]
\label{Fig2} \centering\includegraphics[width=1.0\columnwidth]{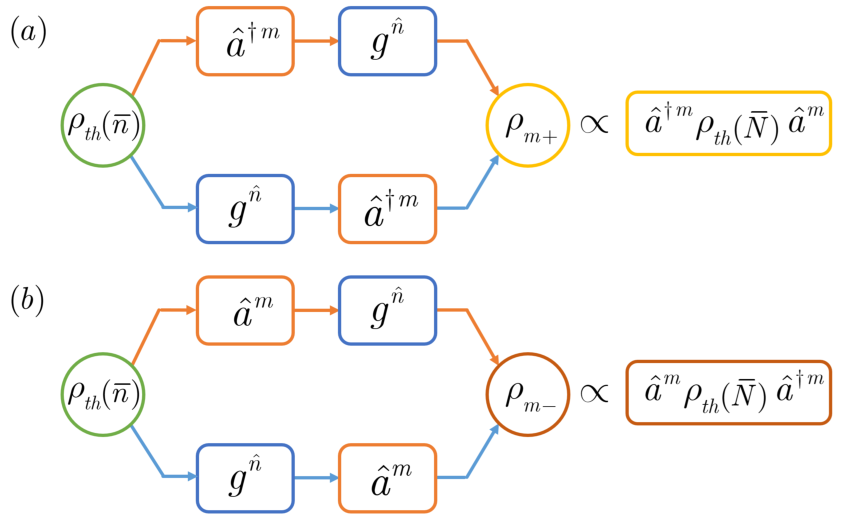}
\caption{(a) Conceptual schemes of generating PAATSs (b) Conceptual schemes
of generating PSATSs.}
\end{figure}

\section{Photon number distributions of considered states}

After analyzing the density matrices for all above states (including TS,
ATS, PAATS and PSATS), we immediately know that their non-diagonal elements
are zero (i.e., $\rho _{kl}=0$\ for $k\neq l$).\ So all these states can be
expanded into the following compound Poisson distribution%
\begin{equation}
\rho =\sum_{k=0}^{\infty }\rho _{kk}\left\vert k\right\rangle \left\langle
k\right\vert ,  \label{b1}
\end{equation}%
where the value of $\rho _{kk}$ denotes the probability of finding the Fock
state $\left\vert k\right\rangle $\ in the state $\rho $. In addition, the $%
\rho _{kk}$s for $\rho _{th}\left( \bar{n}\right) $, $\rho _{th}\left( \bar{N%
}\right) $, $\rho _{m+}$, and $\rho _{m-}$ can be expressed as
\begin{eqnarray}
\rho _{kk\left( \bar{n}\right) }^{\left( 0\right) } &=&\frac{\bar{n}^{k}}{%
\left( \bar{n}+1\right) ^{k+1}},  \notag \\
\rho _{kk\left( \bar{N}\right) }^{\left( 0\right) } &=&\frac{\bar{N}^{k}}{%
\left( \bar{N}+1\right) ^{k+1}},  \notag \\
\rho _{kk}^{\left( m+\right) } &=&\frac{\bar{N}^{\left( k-m\right) }k!}{%
\left( \bar{N}+1\right) ^{k+1}m!\left( k-m\right) !},  \notag \\
\rho _{kk}^{\left( m-\right) } &=&\frac{\bar{N}^{k}\left( m+k\right) !}{%
\left( \bar{N}+1\right) ^{k+m+1}m!k!},  \label{b2}
\end{eqnarray}%
respectively. Therefore, we just need to explore the photon number
distributions (PNDs) for these quantum states.

Taking $\bar{n}=1.5$ with different $g=1.05$, $1.1$, and $1.2$, we plot the
PNDs for $\rho _{0+}$, $\rho _{1+}$, $\rho _{3+}$, and $\rho _{5+}$ in Fig.3
and for $\rho _{0-}$, $\rho _{1-}$, $\rho _{3-}$, and $\rho _{5-}$ in Fig.4.
As the parameter $g$\ increases, the PND of $\rho _{th}\left( \bar{N}\right)
$ become broader (see column\ 1 in Fig.3 and column\ 1 Fig.4). Obviously,
photon-addition or photon-subtraction can change the PNDs of the ATS, but
not simply shift its PNDs. Compared Fig.3 with Fig.4, we find that: (i) the
components including $\left\vert 0\right\rangle $, $\left\vert
1\right\rangle $, $\cdots $, $\left\vert m-1\right\rangle $ are absent for $%
\rho _{m+}$, i.e., $\rho _{kk}^{(m+)}=0$ ($k=0,\cdots ,m-1$); (ii) the PNDs
for $\rho _{m+}$\ and $\rho _{m-}$\ are the same apart from a shift, i.e., $%
\rho _{\left( k+m\right) \left( k+m\right) }^{(m+)}=\rho _{kk}^{(m-)}$. In
other word, the PNDs of\ $\rho _{m+}$ and $\rho _{m-}$ have the same shape,
but $\rho _{kk}^{(m-)}$ starts at $k=0$ and $\rho _{kk}^{(m+)}$ starts at $%
k=m$. In Table I, we list some probabilities for $\rho _{th}\left( \bar{n}%
\right) $, $\rho _{th}\left( \bar{N}\right) $, $\rho _{3+}$, and $\rho _{3-}$
and see $\rho _{00}^{(3+)}=\rho _{11}^{(3+)}=\rho _{22}^{(3+)}=0$, $\rho
_{33}^{(3+)}=\rho _{00}^{(3-)}$, $\rho _{44}^{(3+)}=\rho _{11}^{(3-)}$, and$%
\ \rho _{55}^{(3+)}=\rho _{22}^{(3-)}$.
\begin{figure*}[tbp]
\label{Fig3} \centering\includegraphics[width=2.0\columnwidth]{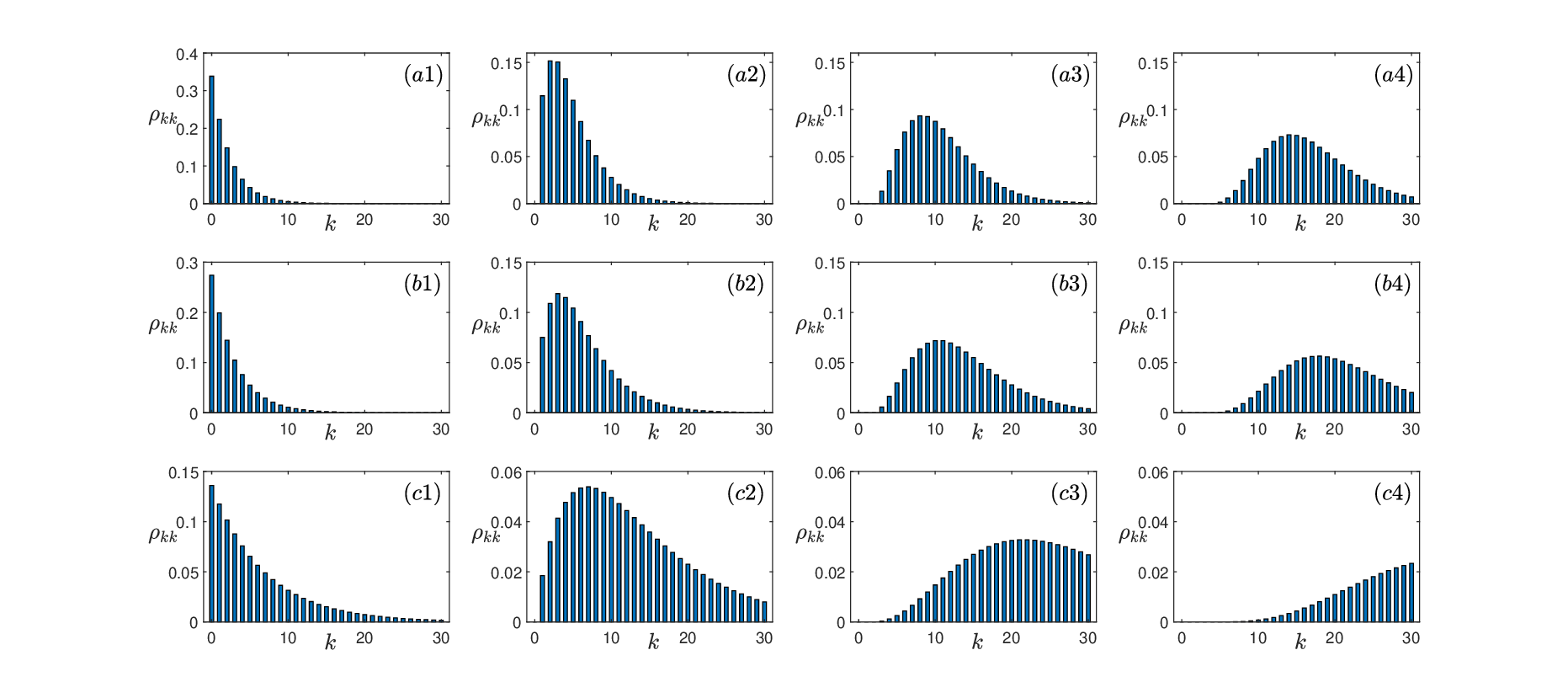}
\caption{PNDs for PAATSs $\protect\rho _{0+}$ (column 1), $\protect\rho %
_{1+} $ (column 2), $\protect\rho _{3+}$ (column 3), $\protect\rho _{5+}$
(column 4) with $g=1.05$ (row 1), $g=1.1$ (row 2), $g=1.2$ (row 3) and fixed
$\bar{n}=1.5$.}
\end{figure*}
\begin{figure*}[tbp]
\label{Fig4} \centering\includegraphics[width=2.0\columnwidth]{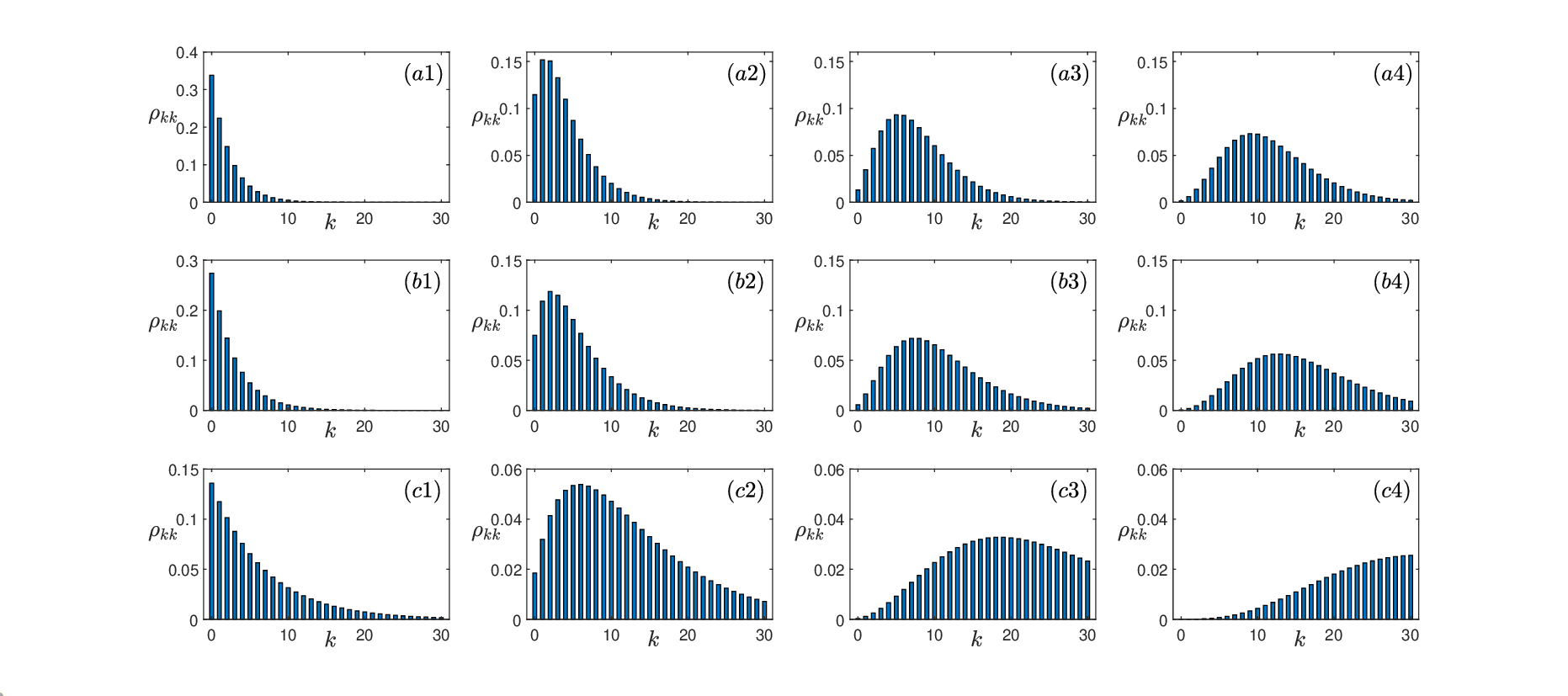}
\caption{PNDs for PSATSs $\protect\rho _{0-}$ (column 1), $\protect\rho %
_{1-} $ (column 2), $\protect\rho _{3-}$ (column 3), $\protect\rho _{5-}$
(column 4) with $g=1.05$ (row 1), $g=1.1$ (row 2), $g=1.2$ (row 3) and fixed
$\bar{n}=1.5$.}
\end{figure*}
\begin{table}[h]
\caption{Some probability values for $\protect\rho _{th}\left( \bar{n}%
\right) $, $\protect\rho _{th}\left( \bar{N}\right) $, $\protect\rho _{3+}$,
and $\protect\rho _{3-}$ with $\bar{n}=1.5$\ and $g=1.2$. Notice: to compare
the values of $\protect\rho _{3+}$\ with those of $\protect\rho _{3-}$.}
\begin{center}
\begin{tabular}{|c||c|c|c|c|}
\hline\hline
PND & $\rho _{th}\left( \bar{n}\right) $ & $\rho _{th}\left( \bar{N}\right) $
& $\rho _{3+}$ & $\rho _{3-}$ \\ \hline\hline
$\rho _{00}$ & $0.4$ & $0.136$ & $0$ & $0.000342102$ \\ \hline
$\rho _{11}$ & $0.24$ & $0.1175$ & $0$ & $0.0011823$ \\ \hline
$\rho _{22}$ & $0.144$ & $0.1015$ & $0$ & $0.00255378$ \\ \hline
$\rho _{33}$ & $0.0864$ & $0.08772$ & $0.000342102$ & $0.00441293$ \\ \hline
$\rho _{44}$ & $0.05184$ & $0.07579$ & $0.0011823$ & $0.00667235$ \\ \hline
$\rho _{55}$ & $0.031104$ & $0.06548$ & $0.00255378$ & $0.00922385$ \\ \hline
$\vdots $ & $\vdots $ & $\vdots $ & $\vdots $ & $\vdots $ \\ \hline
\end{tabular}%
\end{center}
\end{table}

\section{Purities of considered states}

The purity\ is defined as $P=\mathrm{Tr}\left( \rho ^{2}\right) $ for a
quantum state $\rho $. Thus, we can obtain the purity%
\begin{equation}
P_{m+}=\frac{\bar{N}^{2m}}{\left( 2\bar{N}+1\right) ^{2m+1}}\left.
_{2}F_{1}\right. (-m,-m;1,\frac{\left( \bar{N}+1\right) ^{2}}{\bar{N}^{2}})
\label{c2}
\end{equation}%
for $\rho _{m+}$ and the purity%
\begin{equation}
P_{m-}=\frac{\left( \bar{N}+1\right) ^{2m}}{\left( 2\bar{N}+1\right) ^{2m+1}}%
\left. _{2}F_{1}\right. (-m,-m;1,\frac{\bar{N}^{2}}{\left( \bar{N}+1\right)
^{2}})  \label{c3}
\end{equation}%
for $\rho _{m-}$. Here $\left. _{2}F_{1}\right. (a,b;c,d)$\ denotes a
hypergeometric function\cite{40}. In particularly, we have $%
P_{0+}=P_{0-}=1/\left( 2\bar{n}+1\right) $\ for $\rho _{th}\left( \bar{n}%
\right) $ and $P_{0+}=P_{0-}=1/\left( 2\bar{N}+1\right) $ for $\rho
_{th}\left( \bar{N}\right) $.

Figure 5 depicts the purities ($P_{m+}$/$P_{m-}$) as functions of $\bar{N}$\
for different $m$. The bigger $m$ is, the smaller the purities ($P_{m+}$/$%
P_{m-}$)\ are. It was surprising to see that the identity $P_{m+}\equiv
P_{m-}$ is hold for the same $\bar{N}$. Then the question is: why are these
two purities equal? Indeed, we can link $P$ with $\rho _{kk}$ with the rule%
\begin{equation}
P=\sum_{k=0}^{\infty }\rho _{kk}^{2}.  \label{c1}
\end{equation}%
for our considered states. That is to say, the purity for each of these state
is the sum of squares of all its diagonal elements\cite{41}. Thus, we
immediately understand the reason that $P_{m+}$ is always equal to $P_{m-}$
for the same $m$.

\begin{figure}[tbp]
\label{Fig50} \centering\includegraphics[width=0.9\columnwidth]{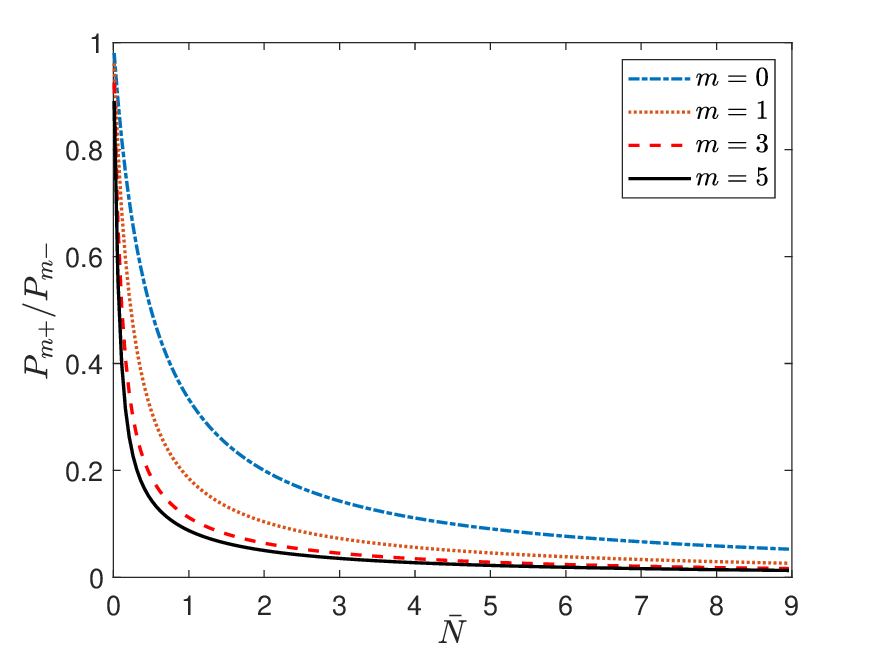}
\caption{Purities ($P_{m+}$/$P_{m-}$) as functions of $\bar{N}$ with $%
m=0,1,3,5$.}
\end{figure}

\section{Wigner functions of considered states}

Gaussian or non-Gaussian state can be judged from its WF. The Wigner
negativity (WN) is an indicator of nonclassicality of quantum state\cite{42}%
. Gaussian state has Gaussian WF without WN. Non-Gaussian state has
non-Gaussian WF with or without WN. The WF for a quantum state $\rho $ is
defined by
\begin{equation}
W_{\rho }\left( \beta \right) =\mathrm{Tr}\left( \rho O_{w}\left( \beta
\right) \right)   \label{d1}
\end{equation}%
with $O_{w}\left( \beta \right) =\frac{2}{\pi }:e^{-2\left( a^{\dag }-\beta
^{\ast }\right) \left( a-\beta \right) }:$ and $\beta =(x+iy)/\sqrt{2}$\cite%
{43}. Corresponding to Fig.3 and Fig.4, we plot their WFs in Fig.6 and
Fig.7, respectively.

WFs for $\rho _{0+}$\ and $\rho _{0-}$ (i.e., $\rho _{th}\left( \bar{N}%
\right) $) have the common Gaussian form with
\begin{equation}
W_{\rho _{th}\left( \bar{N}\right) }\left( \beta \right) =\frac{2}{\pi
\left( 2\bar{N}+1\right) }e^{-\frac{2}{2\bar{N}+1}\left\vert \beta
\right\vert ^{2}}.  \label{d2}
\end{equation}%
which is positive everywhere in the phase space and peaks at the center. As
the factor $g$\ increases, the distribution will become broader, which can
be shown in column 1 in Fig.6 and column 1 in Fig.7. Naturally, the
distribution of the ATS is broader than that of the TS.

WFs for $\rho _{m+}$\ and $\rho _{m-}$\ ($m>0$) have the respective
non-Gaussian forms with%
\begin{eqnarray}
W_{\rho _{m+}}\left( \beta \right) &=&T_{ng}^{(m+)}W_{\rho _{th}\left( \bar{N%
}\right) }\left( \beta \right) ,  \notag \\
W_{\rho _{m-}}\left( \beta \right) &=&T_{ng}^{(m-)}W_{\rho _{th}\left( \bar{N%
}\right) }\left( \beta \right) ,  \label{d3}
\end{eqnarray}%
with respective non-Gaussian terms $T_{ng}^{(m+)}$ and $T_{ng}^{(m-)}$.
Obviously, the PAATS and the PSATS have ring-shaped non-Gaussian
distributions. These results demonstrate that the non-Gaussian operations ($%
\hat{a}^{\dag m}$ and $\hat{a}^{m}$) can render Gaussian states into
non-Gaussian states. From Fig.6, we see that the PAATSs are non-Gaussian
states with WNs. From Fig.7, we see that the PSATSs are non-Gaussian states
without WNs. For example, we have
\begin{eqnarray}
T_{ng}^{(1+)} &=&\frac{4\left( \bar{N}+1\right) }{\left( 2\bar{N}+1\right)
^{2}}\left\vert \beta \right\vert ^{2}-\frac{1}{2\bar{N}+1},  \notag \\
T_{ng}^{(1-)} &=&\frac{4\bar{N}}{\left( 2\bar{N}+1\right) ^{2}}\left\vert
\beta \right\vert ^{2}+\frac{1}{2\bar{N}+1}.  \label{d4}
\end{eqnarray}%
There exists WN for $\rho _{1+}$ if $T_{ng}^{(1+)}<0$, i.e. in the condition
of $\left\vert \beta \right\vert <\sqrt{\left( 2\bar{N}+1\right) /\left(
\bar{N}+1\right) }/2$. But it is impossible to have WN for $\rho _{1-}$
because of $T_{ng}^{(1-)}>0$ for any $\beta $. Similar results are
applicable to cases $m\geq 2$. In order to see WN clearly in Fig.6, we plot $%
W\left( x,y=0\right) $ for $\rho _{m+}$\ as a function of $x$ in Fig.8. From
Fig.8, we see that: 1) the bigger the number $m$\ is, the smaller the WN is;
2) the bigger the proper $g$\ value is, the smaller the WN is. The minimum
WF values are on or below the order of $10^{-2}$, $10^{-3}$, and $10^{-4}$
for cases $m=1,3,5$, respectively.

\begin{figure*}[tbp]
\label{Fig6} \centering\includegraphics[width=1.9\columnwidth]{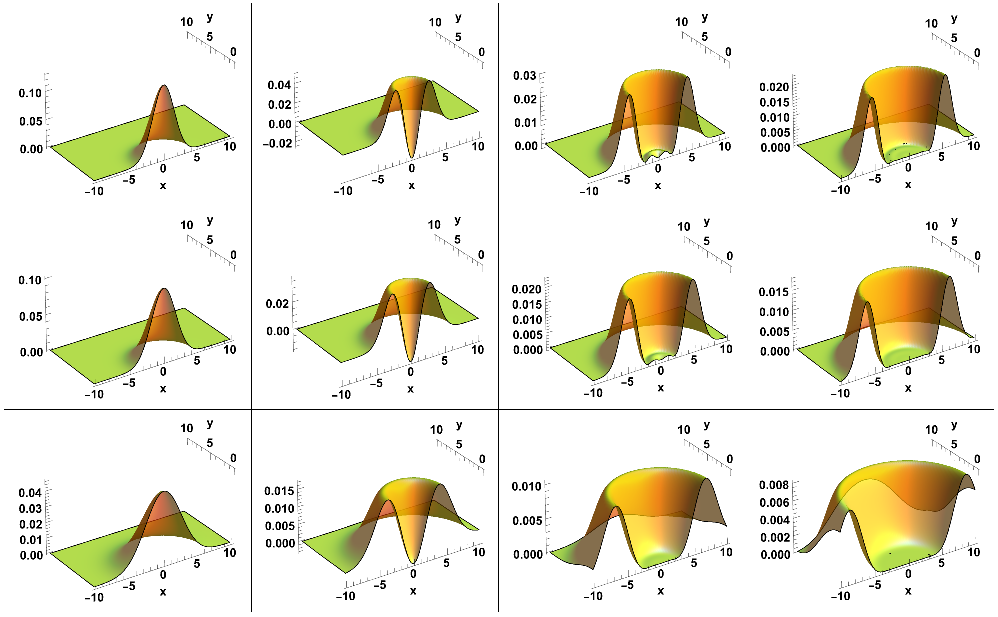}
\caption{Wigner functions of PAATSs $\protect\rho _{0+}$ (column 1), $%
\protect\rho _{1+}$ (column 2), $\protect\rho _{3+}$ (column 3), $\protect%
\rho _{5+}$ (column 4) with $g=1.05$ (row 1), $g=1.1$ (row 2), $g=1.2$ (row
3) and fixed $\bar{n}=1.5$.}
\end{figure*}

\begin{figure*}[tbp]
\label{Fig7} \centering\includegraphics[width=1.9\columnwidth]{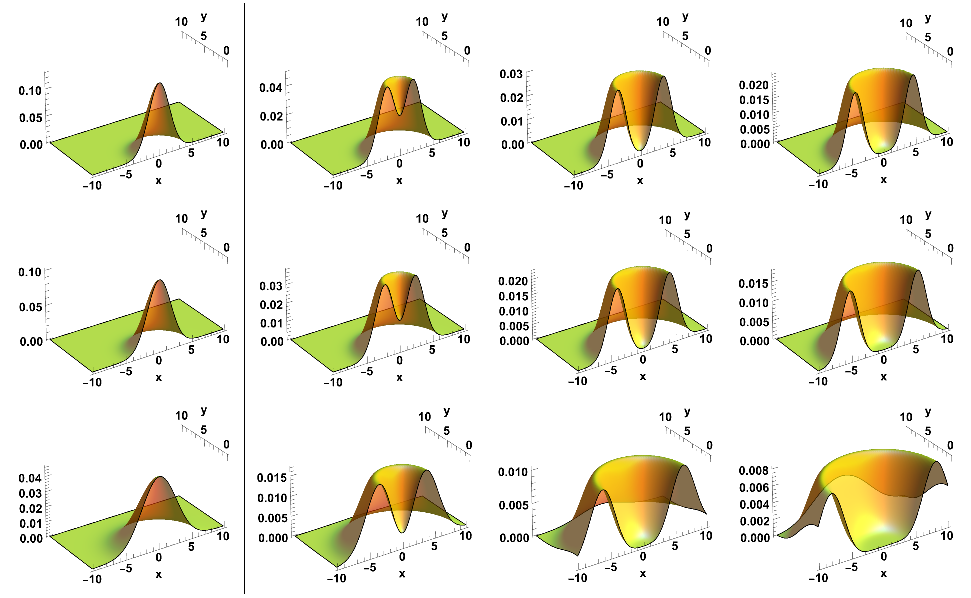}
\caption{Wigner functions of PSATSs $\protect\rho _{0-}$ (column 1), $%
\protect\rho _{1-}$ (column 2), $\protect\rho _{3-}$ (column 3), $\protect%
\rho _{5-}$ (column 4) with $g=1.05$ (row 1), $g=1.1$ (row 2), $g=1.2$ (row
3) and fixed $\bar{n}=1.5$.}
\end{figure*}

\begin{figure*}[tbp]
\label{Fig8} \centering\includegraphics[width=1.9\columnwidth]{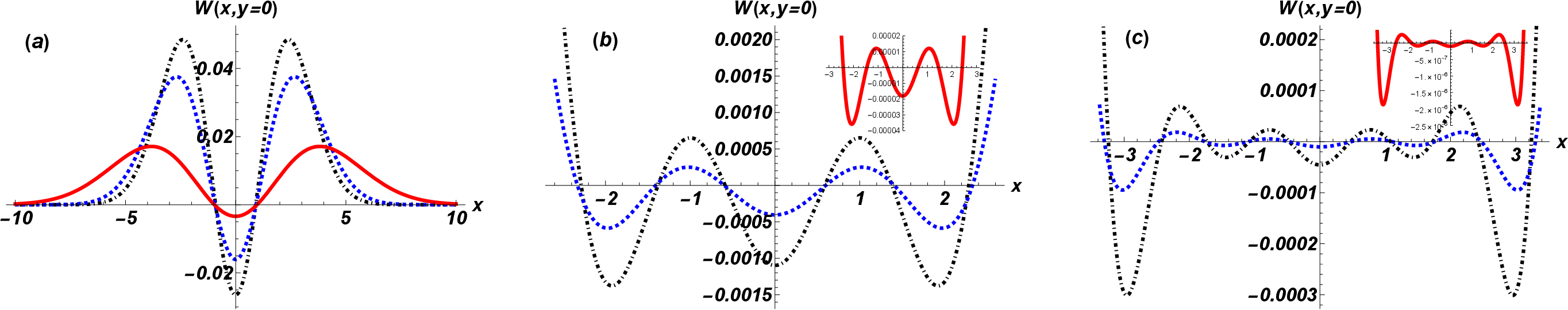}
\caption{Sections for part WFs in Fig.6, for $\protect\rho _{m+}$ with (a) $%
m=1$, (b) $m=3$, (c) $m=5$. For each figure, we take $\bar{n}=1.5$ and $%
g=1.05$ (black dotdashed), $1.1$ (blue dashed), $1.2$ (red solid). The main
character is to show the WNs.}
\end{figure*}

\section{Conclusion}

In summary, we have introduced the ATS by acting $g^{\hat{n}}$ on the TS and
found that the ATS only exist in a certain range of relative parameters
(including $g$\ and $\bar{n}$). Based on the ATS, we have further introduced
the PAATS and the PSATS by operating $\hat{a}^{\dag m}$ and $\hat{a}^{m}$ on
the ATS, respectively. The behaviors of their PNDs, purities, and Wigner
functions are studied by adjusting the interaction parameters ($g$, $\bar{n}$%
\ and $m$).

Our results show that: (1) All our considered states have the compound
Poissonian PNDs. (2) The PAATS and the PSATS with same $m$ have the same
purities. (3) The TS and the ATS are Gaussian states without WNs. (4) The
PAATSs are non-Gaussian states with WNs. (5) The PSATSs are non-Gaussian
states without WNs. Our analysis in the work will be a feasible theoretical
reference for experimenters.

\textbf{Appendix A: Some information of the ATS}

Operating $g^{\hat{n}}$\ on $\rho _{th}\left( \bar{n}\right) $ and using $g^{%
\hat{n}}\left\vert n\right\rangle =g^{n}\left\vert n\right\rangle $, we
obtain the following relation
\begin{equation}
\frac{1}{N_{0}}g^{\hat{n}}\rho _{th}\left( \bar{n}\right) g^{\hat{n}}=\frac{1%
}{N_{0}}\sum_{n=0}^{\infty }\frac{\left( g^{2}\bar{n}\right) ^{n}}{(\bar{n}%
+1)^{n+1}}\left\vert n\right\rangle \left\langle n\right\vert  \tag{A1}
\end{equation}%
with an undetermined coefficient $N_{0}$. In order to form a physical state,
we must ensure $\mathrm{Tr}[\frac{1}{N_{0}}g^{\hat{n}}\rho _{th}\left( \bar{n%
}\right) g^{\hat{n}}]=1$ and then have%
\begin{equation}
\frac{1}{N_{0}(\bar{n}+1)}\sum_{n=0}^{\infty }(\frac{g^{2}\bar{n}}{\bar{n}+1}%
)^{n}=1.  \tag{A2}
\end{equation}%
Obviously, if and only if $g^{2}\bar{n}/(\bar{n}+1)<1$, the series $%
\sum_{n=0}^{\infty }(\dfrac{g^{2}\bar{n}}{\bar{n}+1})^{n}$ is convergent.
Thus, we can obtain $N_{0}=1/[1-\left( g^{2}-1\right) \bar{n}]$ and
reconstruct Eq.(A1) as%
\begin{equation}
\frac{1}{N_{0}}g^{\hat{n}}\rho _{th}\left( \bar{n}\right) g^{\hat{n}}=\frac{1%
}{\bar{N}+1}\sum_{n=0}^{\infty }\dfrac{\bar{N}^{n}}{(\bar{N}+1)^{n}}%
\left\vert n\right\rangle \left\langle n\right\vert  \tag{A3}
\end{equation}%
with $\bar{N}=N_{0}g^{2}\bar{n}$. So the proof of Eq.(\ref{a2}) is completed.

On the other hand, it is impossible to form a quantum state if $g^{2}\bar{n}%
/(\bar{n}+1)>1$ because the series $\sum_{n=0}^{\infty }(\frac{g^{2}\bar{n}}{%
\bar{n}+1})^{n}$ is divergent.

\textbf{Appendix B: Some information of the PAATS and the PSATS}

\textit{(1) Normal ordering forms}

Owe to $\hat{n}\left\vert n\right\rangle =n\left\vert n\right\rangle $, we
know $g^{\hat{n}}\left\vert n\right\rangle =g^{n}\left\vert n\right\rangle $
and $g^{\hat{n}}\left\vert \alpha \right\rangle =e^{(g^{2}-1)\left\vert
\alpha \right\vert ^{2}/2}\left\vert g\alpha \right\rangle $. According to
formula $e^{\lambda \hat{a}^{\dag }\hat{a}}=:e^{(e^{\lambda }-1)\hat{a}%
^{\dag }\hat{a}}:$, we know $g^{\hat{n}}=e^{(\ln g)\hat{a}^{\dag }\hat{a}%
}=:e^{(g-1)\hat{a}^{\dag }\hat{a}}:$ and then obtain

\begin{align}
\rho _{m+}& =\frac{1}{(\bar{N}+1)N_{m+}}\partial _{\mu _{1}}^{m}\partial
_{\nu _{1}}^{m}  \notag \\
& :e^{-\frac{1}{\bar{N}+1}a^{\dag }a+\mu _{1}a^{\dag }+\nu _{1}a}:|_{\mu
_{1}=\nu _{1}=0},  \tag{B1}
\end{align}%
and%
\begin{align}
\rho _{m-}& =\frac{1}{(\bar{N}+1)N_{m-}}\partial _{\mu _{2}}^{m}\partial
_{\nu _{2}}^{m}e^{\mu _{2}\nu _{2}}  \notag \\
& :e^{-\frac{1}{\bar{N}+1}a^{\dag }a+\frac{\bar{N}}{\bar{N}+1}(\mu
_{2}a^{\dag }+\nu _{2}a)}:|_{\mu _{2}=\nu _{2}=0}.  \tag{B2}
\end{align}%
in normal ordering form.

\textit{(2) Density matrix elements}

For state $\rho _{m+}$, we have
\begin{align}
\rho _{kl}^{(m+)}& =\left\langle \left\vert l\right\rangle \left\langle
k\right\vert \right\rangle _{\rho _{m+}}  \notag \\
& =\frac{1}{\left( \bar{N}+1\right) \sqrt{k!l!}N_{m+}}\partial _{\mu
_{1}}^{m}\partial _{\nu _{1}}^{m}\partial _{f}^{k}\partial _{h}^{l}  \notag
\\
& e^{f\mu _{1}+h\nu _{1}+\frac{\bar{N}}{\bar{N}+1}fh}|_{\mu _{1},\nu
_{1},f,h=0}.  \tag{B3}
\end{align}%
For state $\rho _{m-}$, we have%
\begin{align}
\rho _{kl}^{(m-)}& =\left\langle \left\vert l\right\rangle \left\langle
k\right\vert \right\rangle _{\rho _{m-}}  \notag \\
& =\frac{1}{\left( \bar{N}+1\right) \sqrt{k!l!}N_{m-}}\partial _{\mu
_{2}}^{m}\partial _{\nu _{2}}^{m}\partial _{f}^{k}\partial _{h}^{l}  \notag
\\
& e^{\frac{\bar{N}}{\bar{N}+1}\left( \mu _{2}\nu _{2}+\nu _{2}h+\mu
_{2}f+fh\right) }|_{\mu _{2},\nu _{2},f,h=0}.  \tag{B4}
\end{align}

\textit{(3) Wigner functions}

For state $\rho _{m+}$, we have%
\begin{align}
W_{\rho _{m+}}\left( \beta \right) & =\left\langle O_{w}\left( \beta \right)
\right\rangle _{\rho _{m+}}  \notag \\
& =\frac{2}{\pi \left( 2\bar{N}+1\right) N_{m+}}e^{-\frac{2}{2\bar{N}+1}%
\left\vert \beta \right\vert ^{2}}  \notag \\
\partial _{\mu _{1}}^{m}\partial _{\nu _{1}}^{m}& e^{\frac{2\left( \bar{N}%
+1\right) }{2\bar{N}+1}(\beta ^{\ast }\mu _{1}+\beta \nu _{1}-\frac{\mu
_{1}\nu _{1}}{2})}|_{\mu _{1},\nu _{1}=0}.  \tag{B5}
\end{align}%
For state $\rho _{m-}$, we have%
\begin{align}
W_{\rho _{m-}}\left( \beta \right) & =\left\langle O_{w}\left( \beta \right)
\right\rangle _{\rho _{m-}}  \notag \\
& =\frac{2}{\pi \left( 2\bar{N}+1\right) N_{m-}}e^{-\frac{2}{2\bar{N}+1}%
\left\vert \beta \right\vert ^{2}}  \notag \\
& \partial _{\mu _{2}}^{m}\partial _{\nu _{2}}^{m}e^{\frac{2\bar{N}}{2\bar{N}%
+1}(\beta ^{\ast }\mu _{2}+\beta \nu _{2}+\frac{\mu _{2}\nu _{2}}{2})}|_{\mu
_{2},\nu _{2}=0}.  \tag{B6}
\end{align}

\end{document}